\documentclass[final]{IEEEtran}
\usepackage{algorithm}
\usepackage{color}
\usepackage{algpseudocode}
\usepackage{graphics}
\usepackage{epsfig}
\usepackage{cite}
\usepackage{threeparttable}
\usepackage{graphicx}
\usepackage{picinpar}
\usepackage{amssymb}
\usepackage[cmex10]{amsmath}
\usepackage{url}

\usepackage{bm}
\usepackage{setspace}
\usepackage{chngcntr}
\usepackage[
CJKbookmarks=false,
colorlinks,
breaklinks=true,
linkcolor=blue,
anchorcolor=blue,
citecolor=blue,
urlcolor=blue,
]{hyperref}
\usepackage{changes} 

\input epsf

\begin{document}

\title{\huge{New Paradigm for Integrated Sensing and Communication with Rydberg Atomic Receiver}}

\author{Minze Chen, Tianqi Mao, \IEEEmembership{Member,~IEEE}, Yang Zhao, Wei Xiao, Dezhi Zheng, Zhaocheng Wang, \IEEEmembership{Fellow,~IEEE}, Jun Zhang, and Sheng Chen, \IEEEmembership{Life Fellow,~IEEE}

\thanks{Minze Chen, Tianqi Mao, Yang Zhao, Wei Xiao and Dezhi Zheng are with State Key Laboratory of Environment Characteristics and Effects for Near-space, Beijing Institute of Technology. Tianqi Mao is also with Greater Bay Area Innovation Research Institute of BIT. Wei Xiao is also with National Key Laboratory of Science and Technology on Space-Born Intelligent Information Processing. Zhaocheng Wang is with Tsinghua University. Jun Zhang is with the State Key Laboratory of CNS/ATM, Beijing Institute of Technology. Sheng Chen is with Ocean University of China. The corresponding authors are Tianqi Mao and Dezhi Zheng.}%
\vspace*{-5mm}
} %

\maketitle

\begin{abstract}
The RYDberg Atomic Receiver (RYDAR) has been demonstrated to surmount the limitation on both the sensitivity and operating bandwidth of the classical electronic counterpart, which can theoretically detect indiscernible electric signals below $-174$\,dBm/Hz with optical measurement through Rydberg-state atoms. Such miracle has established a new quantum-based paradigm for communications and sensing, which motivates a revolution of the transceiver design philosophies to fully unleash the potential of RYDAR towards next-generation networks. Against this background, this article provides an extensive investigation of Rydberg atomic  communications and sensing from theory to hardware implementations. Specifically, we highlight the great opportunities from the hybridization between the RYDAR and the cutting-edge integrated sensing and communication (ISAC), followed by essential preliminaries of the quantum-based receiver. Then we propose a theoretical framework for broadband ISAC based on RYDAR, demonstrated by the proof-of-concept experiments. Afterwards, the enabling technologies for the ISAC framework are explored ranging from channel characterization, waveform design to array-based receiver configurations, where the open problems are also summarized. Finally, the future applications of RYDAR-based ISAC are envisioned, indicating its significant potential for both civilian and military purposes.
\end{abstract}

\vspace{-3mm}
\section{Introduction}\label{S1} 

The forthcoming sixth-generation (6G) network has been envisioned to realize ubiquitous Internet coverage and pervasive sensing, which motivates the space-air-ground(-sea) integrated network (SAGIN) \cite{Mao_WCM_2022}. By establishing broadband interconnections among heterogeneous terminals with diverse sensing capabilities, SAGIN can empower a plethora of civilian/military applications like Internet of everything, critical communications, maritime reconnaissance, etc. \cite{Geraci_commag_2023}. Nevertheless, such cross-domain heterogeneous network necessitates beyond-line-of-sight communication and radar detection over hundreds of kilometers across the entire electromagnetic spectrum, posing stringent requirements on the sensitivity and detectable bandwidth. This seems to be contradicted with the insurmountable limitation from thermal noise level of $-174$ dBm/Hz in classical fully-electric receiver.

To address this bottleneck, the emerging RYDberg Atomic Receiver (RYDAR), originated from the quantum sensing philosophy, is incorporated to establish a new paradigm for wireless communication receiver design \cite{chen2025harnessing}. Unlike classical dipole antennas that convert incoming electromagnetic waves into modulated currents for measurement, the quantum-based atomic receiver exploits the Rydberg-state atoms and exhibits superior sensitivity to electromagnetic fields, thanks to the exceptionally large electric dipole moments of outermost electrons \cite{Cui_JSAC_25}. The energy-state change, and therefore radio-frequency (RF) irradiation, can be measured by observing intensity of the probe laser that penetrates through alkali Rydberg atoms, exploiting the well-established electromagnetically-induced-transparency Autler-Townes (EIT-AT) phenomenon \cite{schlossberger_rydberg_2024}. Such paradigm-shifting breakthrough, on one hand, circumvents the thermal-noise bottleneck of receiving sensitivity with all-optical readout circuits. On the other hand, diverse Rydberg states can be generated by flexibly manipulating the laser, enabling full-spectrum detection ranging from direct current (DC) to millimeter-wave and terahertz bands \cite{Meyer_PRA_2023}. This breakthrough fundamentally transitions receiver design from classical electronic circuitry to quantum–optical interactions in atomic media, enabling unprecedented sensing‑communication capabilities on a unified platform and motivates a revolution of integrated sensing and communication (ISAC) in next-generation networks \cite{Mao_tcom_22}.

Despite of its attractive merits, the inherent quantum characteristics of the RYDAR have introduced unique challenges in its practical implementation. Specifically, despite of its detectable bandwidth over hundreds of GHz, the RYDAR suffers from limited instantaneous bandwidth below 10\,MHz, mainly attributed to the dephasing mechanism \cite{Bowen_PRA_24}. Detectable bandwidth refers to the total tunable spectral range via different Rydberg transitions, while instantaneous bandwidth is the maximum resolvable signal bandwidth without retuning atomic states. Moreover, classical EIT-AT methods tend to scan a broad range of detuning frequencies for each symbol period, inducing non-negligible latency of {ms-level} that further constrains the sampling rate. These have caused additional difficulties in realizing either high-resolution sensing or high-rate communication towards 6G. Against this background, this article provides an extensive investigation of the RYDAR based ISAC technology. Specifically, preliminaries of RYDAR are introduced from both microscopic and macroscopic perspectives. Afterwards, we propose a novel RYDAR-based ISAC framework that can realize broadband communication and sensing simultaneously, where proof-of-concept demonstrations are presented. Based on the proposed framework, several theoretical breakthroughs for RYDAR are explored, consisting of mathematical channel modeling, waveform design and array-based configurations, focusing on mitigating the bandwidth limitation for 6G applications. Finally, a plethora of potential applications of RYDAR are summarized, unveiling its encouraging prospects for both civilian and military purposes.

\section{Preliminaries of Rydberg Atomic Receivers}\label{S2} 

\textbf{Definition of Rydberg Atoms:} Atoms, the fundamental constituents of matter, consist of a positively charged nucleus surrounded by electrons bound by electromagnetic interactions. In quantum mechanics, the state of an electron in an atom is characterized by three quantum numbers: the principal quantum number $n$, the azimuthal quantum number $l$ and the total angular momentum quantum number $j$. When the outermost electron is excited to a state with a sufficiently large principal quantum number (typically $n > 20$), the atom enters a Rydberg state, which is referred to as Rydberg atom.

\textbf{Characteristics of Rydberg Atoms:}
Due to their large principal quantum numbers $n$, Rydberg atoms exhibit remarkable characteristics. According to the Bohr model, the orbital radius scales as $n^2$, resulting in an enhanced electric dipole moment and sensitivity to external electric fields. Their binding energy decreases with increasing $n$ ($E_b \propto n^{-2}$), meaning they require minimal energy for ionization. Even weak external fields significantly perturb their electronic states, facilitating strong coupling with electromagnetic fields, and underpinning their application in high-sensitivity electromagnetic sensing.

\begin{figure}[t!]
    \centering
    \includegraphics[width=1\linewidth]{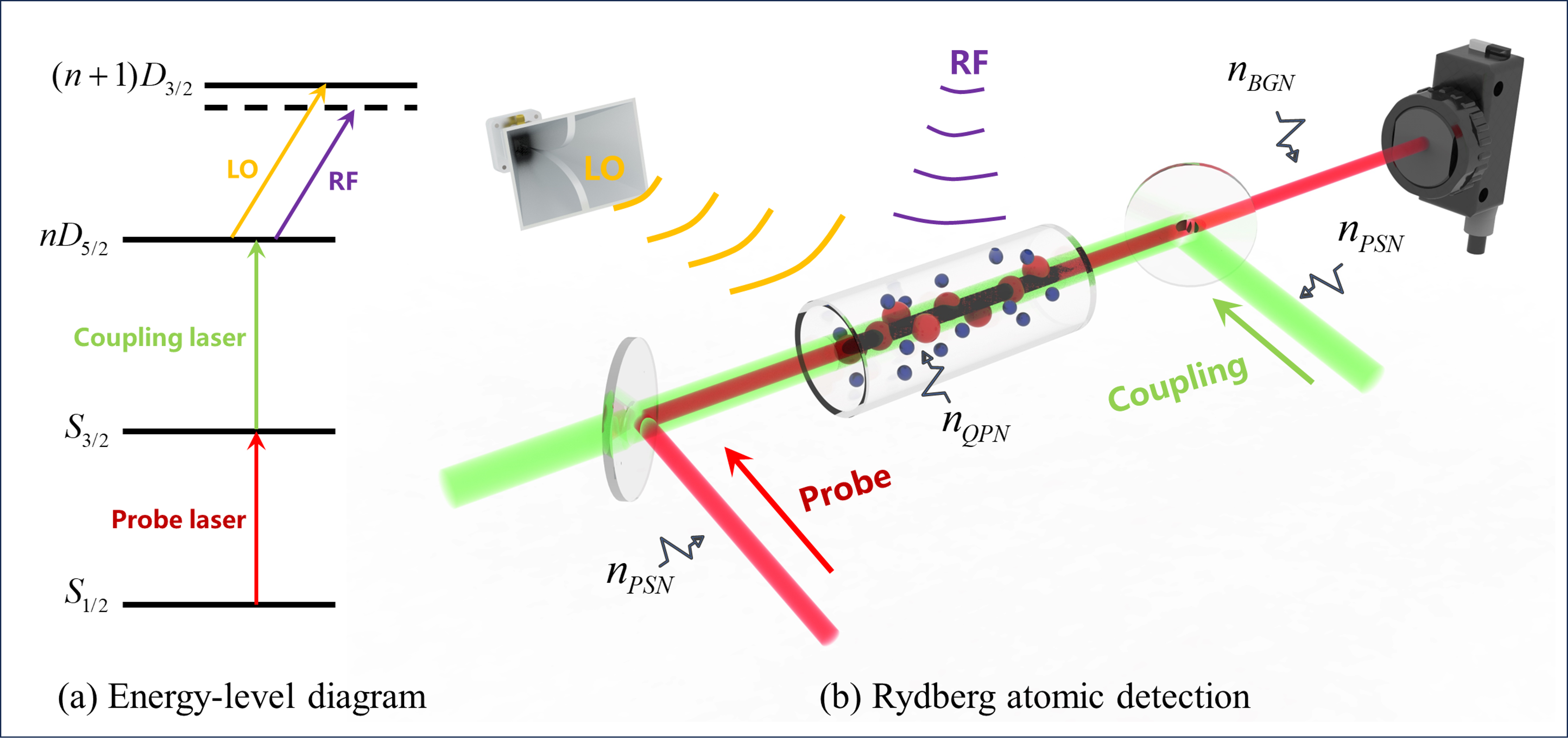}
    \caption{Illustrations of the Rydberg atomic receiving mechanism.}
    \label{fig:mechanism} 
    \vspace{-5mm}
\end{figure}

\textbf{Selection of Rydberg Atoms:}
In selecting Rydberg atoms, Alkali metal atoms are preferred because each has only one outermost electron, which can be easily excited to higher energy levels. Rubidium (Rb) and cesium (Cs) atoms are particularly prevalent, primarily due to their well-characterized energy level structure, relatively long excited-state lifetime, and experimentally accessible excitation conditions.

\textbf{Electronic Transitions of Rydberg Atoms:}
Rydberg atomic systems typically employ a four-level ladder model, excited through a two-photon scheme by two counter-propagating lasers -- the probe and coupling lasers. The probe laser excites electrons from ground state $\left| 1 \right\rangle$ to intermediate state $\left| 2 \right\rangle$, and the coupling laser excites electrons to Rydberg state $\left| 3 \right\rangle$. Atoms in the Rydberg state become sensitive to radio frequency (RF) fields resonating (or near-resonating) with adjacent Rydberg transitions (to state $\left|4\right\rangle$), resulting in AT splitting, exemplifying the Rydberg atom’s response as well as establishing the physical basis for high-sensitivity electromagnetic field detection and precise frequency measurement.

\textbf{Signal Detection:}
Detection of this RF-induced AT splitting employs optical readout via EIT. Initially, the probe laser experiences absorption at resonance (state $\left|2\right\rangle$), but the coupling laser creates a narrow transparency window through quantum interference between states $\left|2\right\rangle$ and $\left|3\right\rangle$, observable as a distinct peak in transmitted probe intensity by a photodetector (PD). When a resonant RF field is applied, this EIT peak splits into two with the splitting interval proportional to the RF field amplitude, quantified by $\Delta f = k\Omega/2\pi = k\mu E/2\pi\hbar$, where $k=\omega_p/\omega_c$ for coupling laser scanning or $k=1$ for probe laser scanning, $\Omega$ is the Rabi frequency, $\mu$ the transition dipole moment between Rydberg states, $E$ the RF field amplitude, and $\hbar$ the reduced Planck constant. Thus, precise measurement of this splitting interval provides high-sensitivity RF signal detection. The working principle of RF signal detection via EIT-AT is illustrated in Fig.~\ref{fig:mechanism}.

\begin{figure}[t!]
    \centering
    \includegraphics[width=1\linewidth]{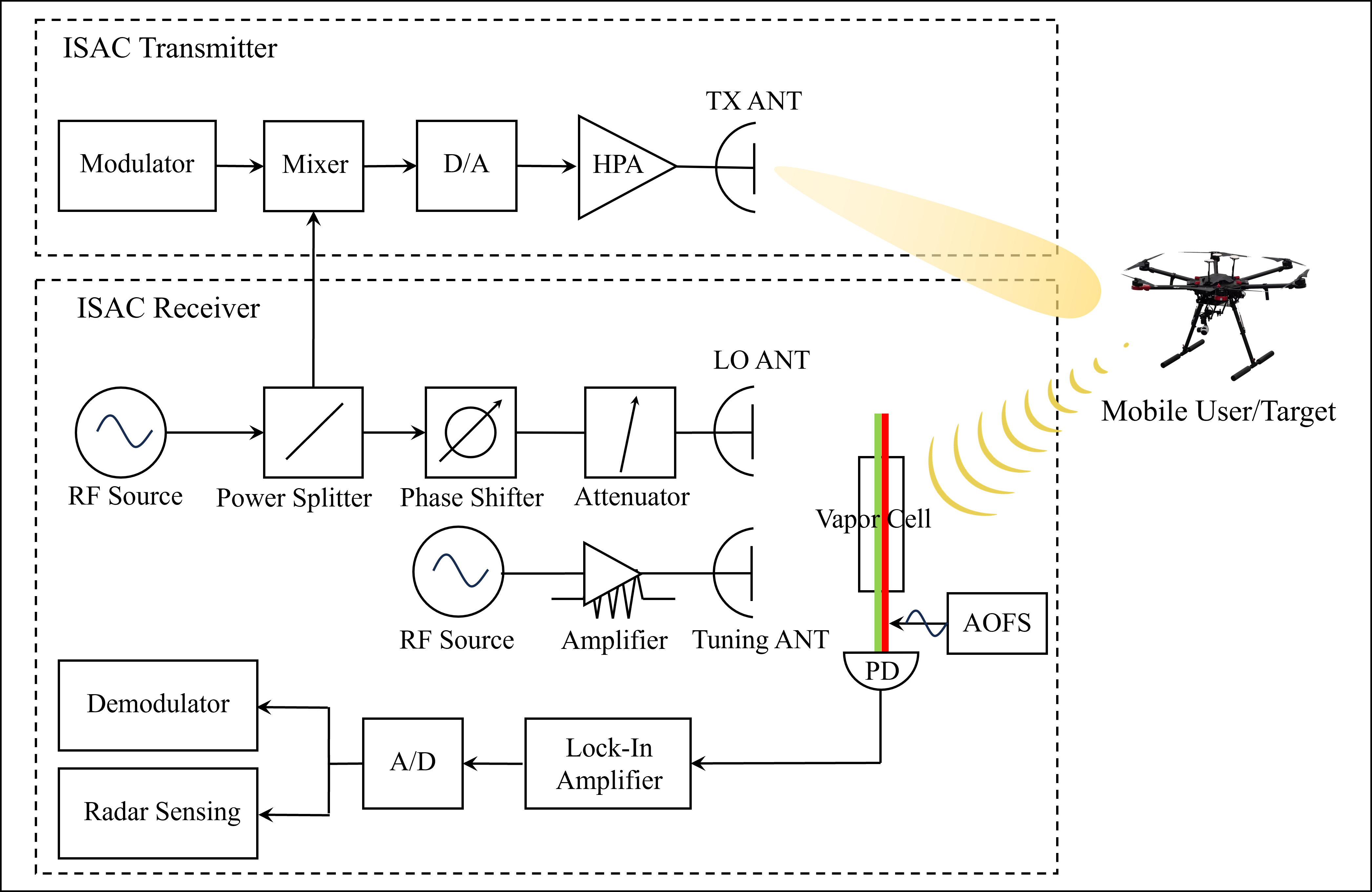} 
    \caption{Transceiver diagram of broadband RYDAR-based ISAC framework.}
    \label{fig:framework} 
    \vspace{-5mm}
\end{figure}

\vspace{-3mm}
\section{Proposed RYDAR-Based ISAC Framework}\label{S3} 

\subsection{Transceiver Principle}\label{S3.1}

As illustrated in Fig. \ref{fig:framework}, the proposed RYDAR-based ISAC framework is exemplified by the monostatic sensing scenario. Herein a RYDAR-empowered base station is established to provide constant data coverage for heterogeneous mobile users like drones and autonomous surface vehicles, whilst concurrently performing radar detection with the communication signals. Specifically, at the transmitter, an unified ISAC waveform is specially designed for simultaneous downlink communication and sensing, aiming at realizing desirable dual-functional performance trade-off. Majority of the examples can be generally classified into communication-centric design and sensing-centric design approaches \cite{Fan_jsac_22}, which highlight the incorporation of sensing function into existing communication waveforms and vice versa, represented by orthogonal frequency division multiplexing (OFDM) and chirp-based signals. Detailed discussions will be presented in Section \ref{S4.2}.

On the other hand, the ISAC receiver generally consists of an all-optical RYDAR structure concatenated by subsequent baseband processing circuits for both communication and sensing. Coherent detection is invoked to distinguish the phase information of electromagnetic waves, making it compatible with amplitude/phase modulation, e.g., phase-shift keying (PSK) and quadrature amplitude modulation (QAM). This necessitates local oscillator (LO) irradiation on the vapor cell, which is also shared with the ISAC transmitter for coherent detection. Under different application scenarios like 6G non-terrestrial networks (NTN), the proposed ISAC receiver is expected to support broadband wireless backhaul and radar sensing in a time-division duplex (TDD) manner, which, however, is hindered by the inherent limitations of typical RYDAR architecture, as elaborated below.

\vspace{-2mm}
\begin{enumerate}
    \item The quantum dephasing mechanism \cite{Bowen_PRA_24} induces insurmountable bottleneck on the instantaneous bandwidth of the RYDAR architecture, i.e., below tens of MHz, which seems to be contradicted with the stringent requirements on next-generation ISAC applications.
    \item Classical EIT-AT scheme attains amplitude detection by directly measuring the AT splitting interval in the spectrum profile of the PD outputs, i.e., the Rabi frequency. This usually necessitates wide-range sweeping of the detuning frequencies, causing ms-level latency within each symbol period that severely degrades the instantaneous bandwidth for communication and sensing.
    \item Conventional EIT-AT measurement primarily depends on the direct-current (DC) component of the probe laser, which is fragile to different noise sources, including the optical background noise and the $1/f$ noise.
\end{enumerate}
\vspace{-2mm}

To circumvent the dephasing imperfection, one feasible solution is to implement frequency-modulated waveforms, e.g., linear-frequency modulated (LFM) and frequency-hopping signals \cite{Cui_sensing_25,Chen_FH_25}, which are capable of mimicking ultra-broadband signals for high-resolution detection regardless of the dephasing phenomenon. To ensure sustained resonance with the frequency-modulated carrier, the system requires dynamic tuning of the laser frequency to adjust the Rydberg atoms' energy level structure in response to discrete frequencies determined by the atomic properties. Furthermore, auxiliary electromagnetic fields must be employed to shift the energy levels via the Stark or Zeeman effect. This approach bridges spectral gaps between discrete frequency points, thereby enabling continuous frequency reception.

\begin{figure}[t!]
    \centering
    \includegraphics[width=1\linewidth]{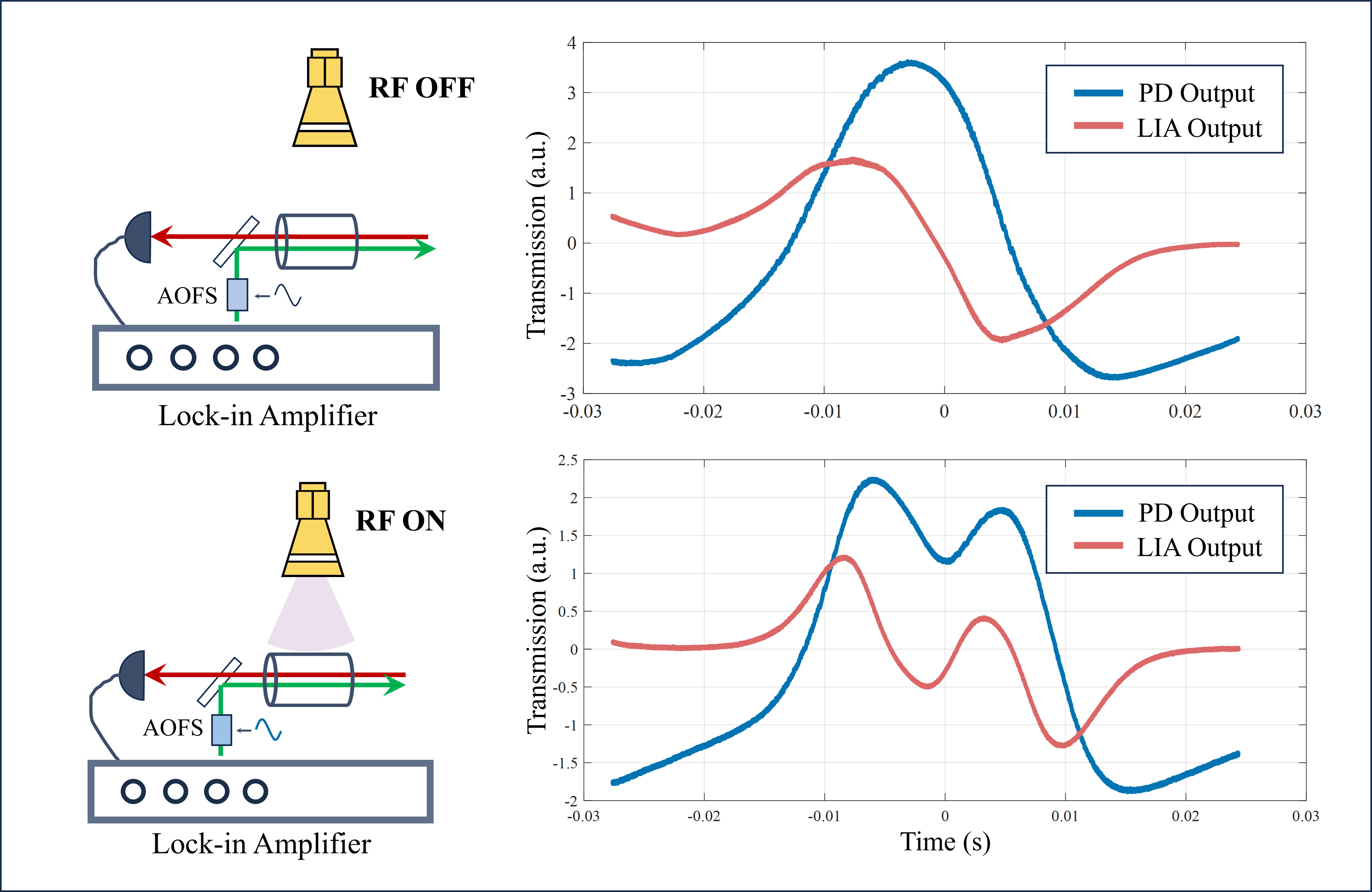}
    \caption{Illustrations of the detected results at the proposed ISAC receiver.}
    \label{fig:LIA} 
    \vspace{-5mm}
\end{figure}

As for the other limitations, an acousto-optic frequency shifter (AOFS) is introduced to generate a small-scale frequency modulation of the coupling laser as shown in Fig.~\ref{fig:framework}, which can significantly eliminate the scanning latency. In this case, the proposed detector keeps monitoring the relative frequency shift of either splitting peak with small-scale frequency scanning during transmission, which no longer measures the splitting interval directly but its dynamic fluctuations for differential calculation of the Rabi frequency. Meanwhile, the PD output is further processed by one lock-in amplifier (LIA), yielding its gradient with respect to the detuning frequency as illustrated in Fig. \ref{fig:LIA}. Let the detuning frequency of the peak corresponding to LO-only irradiation be $f_0$. Asymptotic linearity for the gradient curve in the proximity to $f_0$ can be assumed. Then the relative motion of the peak in subsequent symbol intervals can be proportionally measured from the gradient variation observed at $f_0$, yielding the received Rabi-frequency vector for subsequent demodulation and radar processing. Such detection method can realize high-frequency alternating-current (AC) signal detection by filtering the DC component, leading to enhanced sensitivity against diverse frequency-dependent noise, as validated by existing atomic-physics literature \cite{Budker_Jackson_Kimball_2013}.

\begin{figure*}[htbp]
    \centering
    \includegraphics[width=0.9\textwidth]{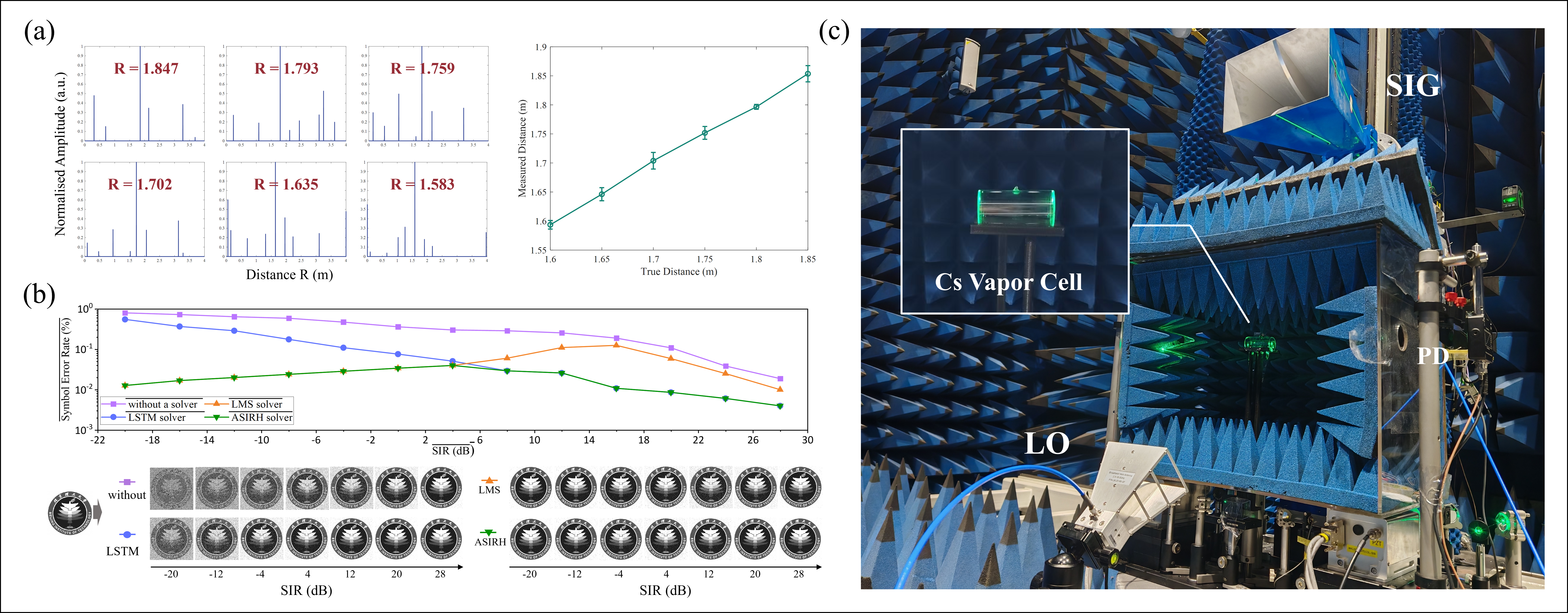}
    \caption{Proof-of-concept: (a)~radar and (b)~communication functionalities. RYDAR in (c) serving as both communication and radar receiver}
    \vspace{-6mm}
    \label{fig:radarandcommunication} 
\end{figure*}

\vspace{-2mm}
\subsection{Experimental Demonstrations}\label{S3.2}

The core of the EIT-AT technology is rubidium or cesium vapor enclosed in a glass cell. In the basic EIT-AT architecture, the system is often used to detect static electric fields. However, signals in practical applications are usually weak and time-varying. To achieve robust capture of such signals, an coherent detection strategy is often employed. The method injects a known LO microwave signal through a horn antenna or waveguide into the atomic gas chamber together with the target signal. Its beat frequency is modulated on the transmitted intensity of the probe laser. A PD extracts this time-varying optical signal, which is equivalent to down-converting the microwave information to a lower frequency or DC for subsequent electronic analysis. It can also use the LO signal as a phase reference to achieve simultaneous recovery of the signal amplitude and phase. As shown in Fig.~\ref{fig:radarandcommunication}\,(c), the core unit consists of a vapor cell, a laser system, microwave feed structures and a PD.

Leveraging this platform, we demonstrate proof-of-concept radar and communication functionalities. In radar mode, the RYDAR detects microwave reflections for distance estimation. Specifically, a modulated microwave signal is transmitted toward a reflective object. The atomic receiver monitors the returning echo. By analyzing the echo's timing delay relative to the transmitted modulation, the system infers the round-trip time of flight and target distance. Laboratory demonstrations (Fig. \ref{fig:radarandcommunication} (a)), although limited to short ranges by experimental constraints, confirm the core radar functionality of the RYDAR. Critically, the same atomic ensemble serves as both communication receiver and microwave radar receiver without hardware reconfiguration -- only signal processing differs. Validations for targets at 1.6-1.9 m demonstrate centimeter-level accuracy (1.04 cm root mean square error) with $\approx$ 15 cm resolution, achieved by analyzing the heterodyne beat note between LO and reflected signal. For communication validation, we developed a hybrid demodulator with network-adaptive filtering \cite{gao2025rydberg}. This system achieved high-performance image transmission using 4-FSK modulation (80 ksyms/s), attaining bit error rate (BER) less than 3.97\%. Remarkably, performance was maintained under significant coexistent wideband radar interference (interference-to-signal ratio (ISR) from -20 dB to -28 dB), as shown in Fig. \ref{fig:radarandcommunication}\,(b). These results underscore RYDARs' inherent versatility as dual-function radar detection nodes and communication terminals, establishing a crucial technical foundation for ISAC systems.

Although the basic application feasibility has been experimentally well verified, further RYDAR still faces challenges, which are summarized as follows.

\textbf{Environmental Sensitivity:} Temperature fluctuations can alter vapor density and atomic resonances, degrading EIT stability. Ambient magnetic fields (via Zeeman and electric fields (via Stark frequency shifts) can perturb the energy levels, thus requiring shielding measures and active compensation. Among these, temperature control is most critical, which can be mitigated by active temperature stabilization combined with optimized cell design to enhance thermal homogeneity. In addition, the accuracy of the optical collimation of the probe and coupling lasers within the vapor cell is critical. 

\textbf{Trade-offs:} The performance of a system is inherently subject to several tradeoffs. The instantaneous bandwidth is limited by the atomic linewidth and coherent detection mechanism. Narrow EIT linewidths are required for high sensitivity, which conflicts with the need for broadband detection. Sensitivity is also constrained by acquisition speed, and detection of weak signals requires longer integration times, which limits the real-time performance of the system. In addition, the simultaneous arrival of multiple signals or the presence of multipath propagation can lead to overlapping AT spectral lines or outlier beat frequencies, generating ambiguities that need to be resolved by complex signal processing algorithms.

\textbf{Path to Practicality:} Beyond sequential transmissions, resolving overlapping signals, which are common in dense scenarios, is essential. Current experimental systems typically rely on discrete optics and large lasers, which faces difficulties in power management and optical path elicitation. Maintaining stable quantum states outside the controlled laboratory environment remains a major challenge. To achieve reliable operation in the field, substantial breakthroughs must be made in terms of structural robustness, size, weight, power consumption and ease of operation. Thus, key scaling priorities include photonic integration for miniaturization and adaptive stabilization against environmental perturbations.

\section{Breakthrough of RYDAR-Based ISAC}\label{S4}

\subsection{Mathematical Channel Modeling}\label{S4.1}

To facilitate subsequent researches on transceiver algorithms, it is essential to constitute a mathematical channel model that can fully capture its quantum peculiarities, for the proposed RYDAR-based ISAC framework. In this section, a monostatic RYDAR structure with homodyne detection is considered under high-mobility scenarios as shown in Fig. \ref{fig:framework}. Herein the overall equivalent channel can be characterized by cascading the air-interface and atomic-reception segments, which will be extensively discussed as below.

\textbf{Atomic Air Interface:} This segment characterizes the electromagnetic propagation from mobile users/targets to the atomic vapor cell, which resembles classical wireless ISAC systems. Classical ISAC systems usually assume time-frequency doubly dispersive channel for communications and flat-fading channel for round-trip sensing, which is no longer applicable for the RYDAR architecture. As discovered in \cite{Cui_sensing_25}, the received signal-to-noise ratio (SNR) suffers from pronounced selectivity across different detuning frequencies, which, therefore, should be incorporated into the end-to-end channel for both communication and sensing. 

\textbf{Atomic Reception:} This segment involves the sophisticated interactions between Rydberg atoms and incident electromagnetic signals, and diverse imperfections are induced by the RYDAR architecture. As described in Section~\ref{S3.1}, an LO irradiation that coherently interacts with the incident RF signals is employed for amplitude and phase detection. For brevity, we consider the received signals from one arbitrary path. Denote the RF signal and LO component as $E_\text{M}=A_\text{M}\cos(\omega_\text{M}t + \phi_\text{M})$ with $M\in\{\text{LO},\text{RF}\}$, respectively, where $\omega_\text{LO}=\omega_\text{RF}$ for homodyne detection. Both amplitude and phase information of the communication/sensing signals $E_\text{RF}$ can be recovered from the in-phase/quadrature coherent detection. Below we only present the in-phase branch for brevity. By setting ${A_{LO}}\gg {A_{RF}}$, the superposition of the two components can be approximated as $E_{\text{sum}} = \left({A_\text{LO}} + {A_\text{RF}}\cos (\Delta \phi)\right)\cos ({\omega _\text{LO}}t + {\phi_\text{LO}})$, where $\Delta \phi=\phi_\text{RF}-\phi_\text{LO}$. The envelope of $E_{\text{sum}}$, i.e., $\left({A_\text{LO}} + {A_\text{RF}}\cos (\Delta \phi)\right)$, can then be obtained through the EIT-AT measurement of Rabi frequency.

Unlike traditional EIT-AT methods, the proposed framework invokes small-scale frequency modulation onto the coupling laser rather than mandatory wideband frequency sweeping, whose optical output is filtered by LIA for sensitivity enhancement. Specifically, the Rabi frequency can be differentially calculated from the observed gradient variation at $f_0$ by exploiting the asymptotic linearity for the gradient curve. Following this strategy, the observed gradient at $f_0$, i.e., the LIA outputs for subsequent processing, can be derived as
\begin{align}\label{eq1}
  & y = {\left. {\frac{{{\partial ^2}{P_\text{out}}}}{{\partial {f}^2}}} \right|_{{f}= {f_0}}}\frac{1}{{2\pi }}\frac{\mu }{\hbar }{A_\text{RF}}\cos (\Delta \phi ) \nonumber \\
	& = \frac{{\Omega _p^2}}{{(2\Omega _p^2\! +\! \Omega _c^2){^2}}}\frac{1}{{2\pi }}\frac{\mu }{\hbar }{A_\text{RF}}\cos (\Delta \phi )\! =\! \frac{k_0}{{2\pi }}\frac{\mu }{\hbar }{A_\text{RF}}\cos (\Delta \phi ), \!  
\end{align}
where $P_\text{out}$ stands for the PD output, $f$ denotes the detuning frequence of coupling laser, and $k_0=\frac{{\Omega _p^2}}{{(2\Omega _p^2 + \Omega _c^2){^2}}}$ denotes the slope of LIA outputs within the linear region, while $\Omega _p$ and $\Omega _c$ are constants that are proportional to the optical power of probe and coupling lasers, respectively. 

Despite the immunity against thermal noise, the peculiar RYDAR structure has induced diverse extrinsic/intrinsic noise components that contaminate the LIA outputs:
\begin{enumerate}
    \item Extrinsic noise defines the background noise components originated from the external interferences and environmental blackbody radiation, which can be approximately modeled as an additive white Gaussian noise (AWGN) to the incident RF signals.
    \item Intrinsic noise generally involves the quantum projection noise and the photon shot noise. The former arises from the fundamental statistical fluctuations inherent to the quantum measurements, which characterizes the inevitable uncertainty in determining energy transition probabilities. On the other hand, photon shot noise results from random perturbation of the coupling/probe lasers, which can distort the PD outputs, especially the dual-peak shape in Fig. \ref{fig:LIA}. Therefore, such perturbation causes severe deviations for the slope parameter $k_0$, which yields multiplicative noise components, making subsequent signal processing rather challenging.
\end{enumerate}

By incorporating the aforementioned noise components, the input-output relationship of the atomic reception segment can be reformulated as
\begin{equation}
y = \frac{1}{{2\pi }}({k_0} + {n_\text{PSN}})\frac{{{\mu}}}{\hbar }({A_\text{RF}}\cos (\Delta \phi ) + {n_\text{BGN}}+{n_\text{QPN}}),
\end{equation}
where $n_\text{PSN}$ denotes the equivalent photon shot noise, $n_\text{BGN}$ and $n_\text{QPN}$ represent the background and  quantum projection noises, respectively. Empirically, $n_\text{PSN}$, $n_\text{BGN}$ and $n_\text{QPN}$ can be approximated as Gaussian variables \cite{Cui_JSAC_25}.

\vspace{-3mm}
\subsection{Broadband Waveform Design}\label{S4.2}   

As discussed in Section~\ref{S3.1}, main-stream ISAC waveforms can be roughly classified into communication-centric and sensing-centric designs. Communication-centric waveform design, represented by frequency-division multiplexing (FDM), is still immature for practical RYDAR architecture, due to the intrinsic instantaneous bandwidth limitation and intractable demodulation \cite{liu_nc_2022}. Such issues can be naturally circumvented by typical sensing-centric design like LFM signals \cite{Cui_sensing_25,Chen_FH_25}. In \cite{Cui_sensing_25}, $50$-MHz equivalent baseband bandwidth was attained with time-varying LFM waveform. Specifically, the transmit power was dynamically manipulated according to the received SNR selectivity across different detuning frequencies, which mitigated the sensitivity loss induced by the inevitable ``deep fading'' within the spectrum. Despite of the superiority as validated numerically, its practical implementation is still obstructed by excessive complexity and unsatisfactory sensitivity level. These difficulties can be addressed by employing the frequency-hopping signals for stepped-frequency synthesis, which has been demonstrated to attain a synthesized bandwidth exceeding 1\,GHz \cite{Chen_FH_25}. The aforementioned literature mainly concentrate on RYDAR-based target sensing, where ISAC waveform design is still at nascent stages. One straightforward strategy is data transmission over frequency-modulated waveform with spread-spectrum philosophy like PSK-LFM signals, which, however, faces peculiar challenges for dual-functional trade-off. For instance, compared with \cite{Cui_sensing_25}, the power control optimization can be more sophisticated by involving multiple target functions and incorporating the time-frequency selectivity of propagation channel fading, rather than the intrinsic counterpart of the RYDAR sensitivity.

\vspace{-3mm}
\subsection{Array-Based Configurations}\label{S4.3}

Extending classical single-RYDAR structure to the array-based configuration can definitely unleash its potential for ISAC applications. On one hand, the array-based deployment allows angle-of-arrival estimation, thus enhancing the spatial awareness critical for radar/localization tasks. On the other hand, the array-based RYDAR can attain desirable spatial diversity/multiplexing gains for improved system capacity and sensitivity. The array-based RYDAR typically consists of vapor-cell elements arranged in linear/planar configurations. Each element, penetrated by intersecting laser beams, can scaled down to sub-wavelength (tens to hundreds of $\mu \mathrm{m}$), enabling extremely compact arrays with precise spatial resolution. The array architecture can be generally classified into centralized and distributed schemes.

In a \textit{centralized design}, a single pair of probe and coupling lasers is split among multiple vapor cells, ensuring coherent optical excitation. Optical signals from each cell are then collected onto multi-channel PDs or imaging detectors. This approach maintains a common optical frequency reference but demands highly precise optical alignment and stable beam distribution across all cells.

In a \textit{distributed design}, each vapor cell has dedicated lasers and photonic detection, operating as an independent node. While this increases hardware complexity and requires individual optical alignment per cell, it improves robustness by eliminating single points of failure inherent in shared optical paths and simplifies tolerance requirements for each unit.

Future monolithic integration of Rydberg arrays onto photonic chips -- leveraging integrated optical waveguides, micro-fabricated atomic vapor microcells/microcavities, and potentially on-chip detectors -- could enable mass production, drastically reduce size and power consumption, and significantly enhance scalability, mirroring.

\begin{figure*}[t!]
    \centering
    \includegraphics[width=0.6\linewidth]{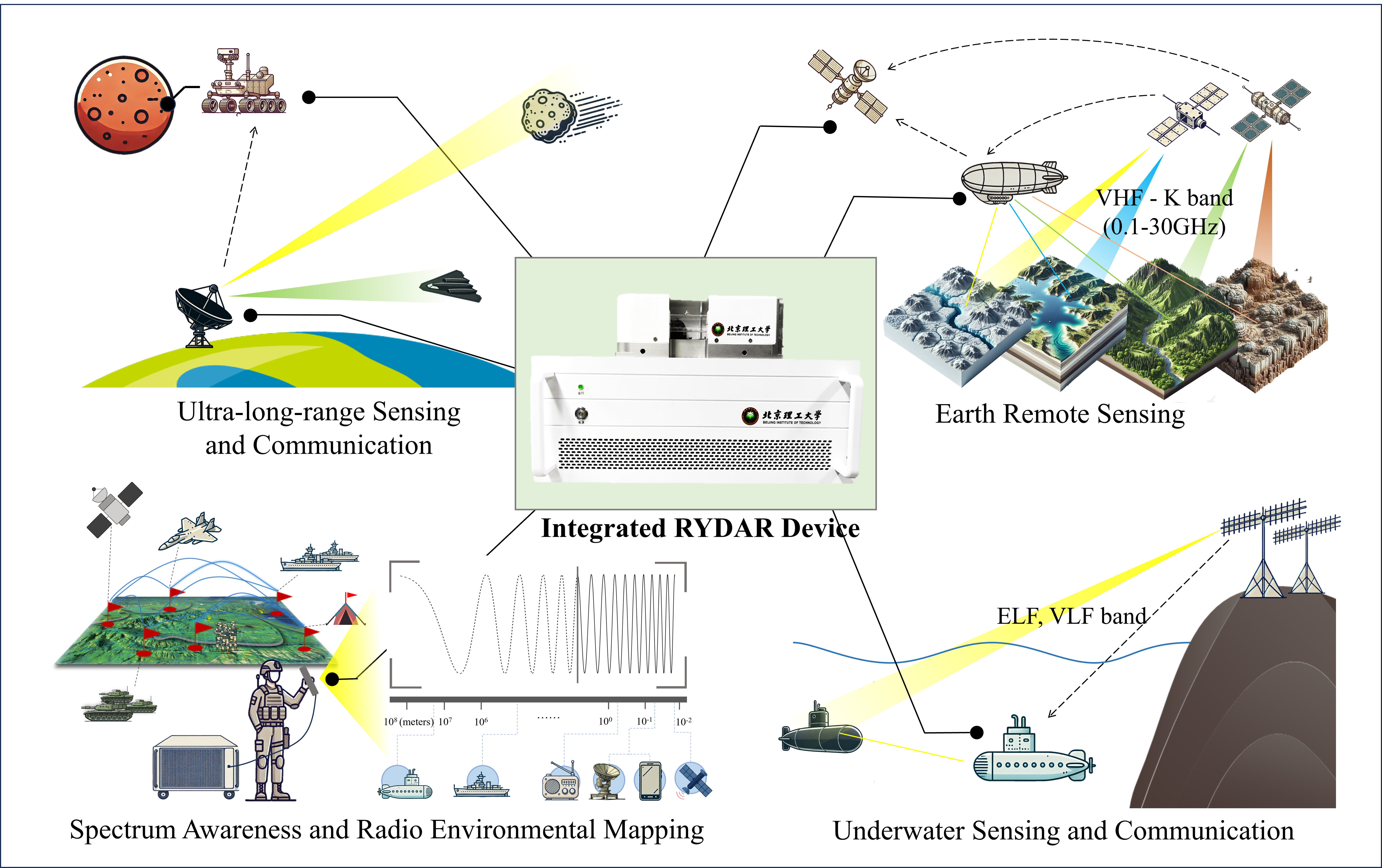}
    \caption{Typical applications for RYDAR-based ISAC.}
    \label{fig:applications}
    \vspace{-5mm}
\end{figure*}

\textbf{Phase Coherence and Timing:} The consistency of the system phase and timing is critical to enable effective multiple-input multiple-output (MIMO) operation (e.g., beam forming or coherent signal processing). When constructing a multi-unit Rydberg sensing array, it is necessary to ensure that the probe/coupling lasers of each unit remain phase-locked both spatially and temporally, so that the atomic responses of the channels are comparable in phase. Meanwhile, when downconverting RF signals for detection, it is necessary for the channels to share the same microwave LO or to maintain phase synchronization between the LOs through a phase-locked loop, so that the downconverted baseband signals can be coherently synthesized. The energy-level coherence bandwidths or decay rates on the Rydberg state (typically a few kHz to a dozen kHz) and the corresponding group response times (on the order of tens of microseconds) must also be considered in the synchronization design, as they determine the limits of the atomic response to fast modulated or pulsed signals.

\textbf{AI-Assisted Calibration and Control:} The calibration and tuning of large-scale atomic arrays not only require accurate calculations of a large number of environmental and atomic parameters to demodulate the signal, but also demand detailed calibration of the Rydberg energy-level structure for fine-tuning. Machine learning (ML) and artificial intelligence (AI) techniques provide a powerful solution to these challenges. AI-driven algorithms can robustly tune the RYDAR system to maintain high-sensitivity reception. This includes using an automated real-time feedback control loop to dynamically compensate for system drift, such as variations in vapor-cell temperature or laser intensity. Additionally, array calibration is performed automatically using known reference signals to adaptively correct phase and amplitude mismatches among units. The noise-robust demodulator in \cite{liu_nc_2022} provides a promising pathway for such array calibration. Such adaptive learning mechanisms are critical for handling the quantum-mechanical phenomena, nonlinear behaviors, and noise characteristics inherent in large-scale Rydberg arrays, and are expected to significantly outperform traditional heuristic tuning methods. 

\section{Potential Applications of RYDAR-based ISAC}\label{S5} 

As shown in Fig.~\ref{fig:applications}, RYDAR-based ISAC is envisioned to empower a plethora of unprecedented civilian/military applications, which will be exemplified as follows.

\vspace{-3mm}
\subsection{Ultra-Long-Range Sensing and Communication}\label{S5.1}
There have been persistent efforts on the RYDAR technology to approach its exceptional theoretical quantum limit. For instance, experiments have demonstrated that RYDAR can approach the sensitivity limit of dipole antennas at specific microwave frequencies (e.g., $13.9$ GHz) \cite{schlossberger_rydberg_2024}.
Such progress will definitely revolutionize the ultra-long-range ISAC applications, e.g., inter-planetary transmission, deep-space exploration and stealthy platform detection, especially empowered by the breakthrough on the array-based RYDAR. 

\vspace{-3mm}
\subsection{Earth Remote Sensing}\label{S5.2}

Conventional Earth remote sensing requires multiple radar sensors operating across distinct bands (e.g., Very High Frequency (VHF) for subsurface imaging, K-band for precipitation monitoring). RYDAR uniquely provides wideband coverage across these frequencies, consolidating functionalities of multiple narrowband sensors. This reduces payload size, weight, and cost for airborne/spaceborne platforms. Crucially, their post-deployment frequency reconfigurability enables dynamic adjustments of sensing frequencies and parameters -- a decisive advantage over fixed-frequency hardware.

\vspace{-3mm}
\subsection{Radio Environmental Mapping}\label{S5.3}

RYDAR is highly competitive in real-time spatial mapping of the spectrum occupancy, signal strength and interference distribution. Specifically, traditional collection of multi-functional receivers at different frequencies can be readily replaced by one single RYDAR with full-spectrum monitoring. Furthermore, the RYDAR presents superior performances especially under sophisticated radio environments, e.g., the city canyon and battle field, thanks to its non-metallic material highly robust to strong electromagnetic jamming. Current researches are mainly focusing on scenario-specific validation (e.g., military-base protection). Open issues like limited spectrum sweeping speed and inconsistent sensitivity level for continuous spectrum scanning, still require investigation \cite{borowka2024continuous}.

\vspace{-3mm}
\subsection{Underwater Sensing and Communication}\label{S5.4}

The frequency bands for underwater electromagnetic communications are in the Extremely Low Frequency (ELF: 3-30 Hz) or Very Low Frequency (VLF: 3-30 kHz), where traditionally hundreds of meters of antennas are required for reception. Highly sensitive reception of these signals can be achieved with centimeter-scale Rydberg atomic vapor cells. In practice, the quantities being measured are periodic frequency shifts, not AT splits. In the long term, with technological breakthroughs, miniaturized RYDAR are expected to be integrated into autonomous underwater vehicles (AUVs) or submarines for covert communications, high-precision environmental sensing, and navigational aids, thus changing the landscape of underwater detection and communications.

\section{Conclusion}\label{S6}

This work has provided a comprehensive investigation of the RYDAR for ISAC, covering fundamental principles, theoretical modeling, hardware proof-of-concept, and envisioned applications. Notably, compared with conventional RF receivers, RYDARs offer three key advantages: exceptional sensitivity, ultra-broadband spectral coverage, and a highly compact form factor. Their dual-function sensing/communication capabilities have been experimentally validated, confirming the practicality of an all-optical, broadband ISAC receiver. The main enabling advances demonstrated here include detailed channel modeling, tailored waveform design, and array-based configurations. Looking ahead, RYDAR-based systems are expected to enhance future applications in areas such as ultra-long-range communication links, remote sensing, spectrum awareness, and underwater communications. Although challenges such as environmental robustness remain, this quantum-enabled paradigm paves a revolutionary path toward 6G-and-beyond networks, seamlessly integrating sensing and communication on a unified atomic platform.

\vspace{-3mm}
\section*{Acknowledgment}

This work was supported in part by Beijing Outstanding Young Scientist Program under Grant JWZQ20240101014, in part by the Open Project Program of State Key Laboratory of CNS/ATM under Grant 2024B12, in part by Young Elite Scientists Sponsorship Program by CAST under Grant 2022QNRC001, in part by National Key Laboratory of Science and Technology on Space-Born Intelligent Information Processing under Grant TJ-03-24-04 and in part by Beijing Institute of Technology Research Fund Program for High level talents under Grant RCPT-6120230097.

\vspace{-3mm}
\bibliographystyle{IEEEtran} 
\bibliography{ref} 

\vspace{-12mm}
\begin{IEEEbiographynophoto}{Minze Chen} is pursuing a Ph.D. degree in the School of Interdisciplinary Science, Beijing Institute of Technology, Beijing, China. His research focuses on quantum sensing, atomic sensors, remote sensing and Rydberg physics.
\end{IEEEbiographynophoto}
\vspace{-12mm}
\begin{IEEEbiographynophoto}{Tianqi Mao} [M] is currently an Associate Professor with the School of Interdisciplinary Science, Beijing Institute of Technology. His research interests include modulation, waveform design and signal processing for wireless communications, integrated sensing and communication, terahertz communications, and visible light communications. 
\end{IEEEbiographynophoto}
\vspace{-12mm}
\begin{IEEEbiographynophoto}{Yang Zhao} is currently pursuing a Master's degree in the School of Interdisciplinary Science, Beijing Institute of Technology.
\end{IEEEbiographynophoto}
\vspace{-10mm}
\begin{IEEEbiographynophoto}{Wei Xiao} is currently an Assistant Professor at the School of Interdisciplinary Science, Beijing Institute of Technology, Beijing, China. His research focuses on atomic, molecular, and optical physics, with a particular emphasis on quantum sensing and its practical applications.
\end{IEEEbiographynophoto}
\vspace{-12mm}
\begin{IEEEbiographynophoto}{Dezhi Zheng} is currently a Professor with the School of Interdisciplinary Science, Beijing Institute of Technology, Beijing, China. His main research interests include airborne information detection, extreme signal measurement technology, and sensor sensitivity mechanisms.
\end{IEEEbiographynophoto}
\vspace{-12mm}
\begin{IEEEbiographynophoto}{Zhaocheng Wang} [F] is currently a Professor at the Department of Electronic Engineering, Tsinghua University. He was a recipient of IEEE Scott Helt Memorial Award, IET Premium Award, IEEE ComSoc Asia-Pacific Outstanding Paper Award, and IEEE ComSoc Leonard G. Abraham Prize.
\end{IEEEbiographynophoto}
\vspace{-12mm}
\begin{IEEEbiographynophoto}{Jun Zhang} is a Professor with the Beijing Institute of Technology and also the Secretary for the Party Committee of the Beijing Institute of Technology. He is a member of the Chinese Academy of Engineering. His research interests are networked and collaborative air traffic management systems, covering signal processing, integrated and heterogeneous networks, and wireless communications.
\end{IEEEbiographynophoto}
\vspace{-12mm}
\begin{IEEEbiographynophoto}{Sheng Chen} [LF] is a Professor with the School of Electronics and Computer Science, University of Southampton, U.K. He is also currently a Distinguished Professor with Ocean University of China. He is a Fellow of the Royal Academy of Engineering (FREng), of the Asia-Pacific Artificial Intelligence Association, and of the IET.
\end{IEEEbiographynophoto}

\end{document}